\documentstyle{laa}
\input epsf
\def\msun{$M_{\odot}$}
\def\msunyr{\msun\ ${\rm yr}^{-1}$}
\def\lsun{$L_\odot$}

\begin{document}

   \thesaurus{06         
              (03.13.2; 08.02.5; 08.06.3; 08.14.2)}

   \title{Recurrent Novae at Quiescence: Systems with Giant Secondaries}
   
   \author{G.C. Anupama \inst{1} \and J. Miko{\l}ajewska \inst{2}}
   
   \offprints{G.C. Anupama}
   
   \institute{Indian Institute of Astrophysics, Bangalore 560034, India
   \and
    Nicolaus Copernicus Astronomical Center, Bartycka 18, 00-716 Warsaw, Poland}

   \date{Received August 24, 1998; accepted December 9, 1998}

   \maketitle

   \begin{abstract}
   Spectroscopic and photometric behaviour of the class of recurrent novae with
   giant secondaries (T Coronae Borealis, RS Ophiuchi, V3890 Sagittarii and
   V745 Scorpii) at quiescence are presented in this study. The hot component
   in these systems is variable, with the variability manifesting as variability
   in the ultraviolet luminosity, the ultraviolet and optical emission line fluxes and
   in the $UBV$/visual magnitudes. The variations are uncorrelated with the binary
   orbital motion. The observed ultraviolet+optical spectral characteristics of
   the hot component in these systems can be explained by a white dwarf+accretion
   disc embedded in an envelope of wind from the M giant secondary. We suggest
   the observed variations are a result of (a) fluctuations in the mass accretion
   rate; (b) changes in the column density of the absorbing wind envelope, which
   is optically thick.
      
      
   \end{abstract}

\section{Introduction}

The recurrent novae (RNe) form a small, but heterogeneous group of cataclysmic
variable stars that undergo classical nova-like outbursts, reaching luminosities
$M_V\le -5.5$, at intervals of the order of decades. The nova outbursts are 
accompanied by the ejection of matter at velocities 
$V_{\rm{exp}}\ge 300$~km~s$^{-1}$. Some RNe (U Scorpii, V394 Coronae Australis,
LMC 1990 No.\ 2, T Pyxidis) are short period binaries similar to the classical
novae and consist of a white dwarf and a dwarf secondary. In contrast, the
other members of this group (RS Ophiuchi, T Coronae Borealis, V745 Scorpii, V3890
Sagittarii) are long
period binaries, with periods of the order of several hundred days, and
consist of a hot white dwarf and a red giant like the symbiotic binary systems.
The nova outbursts in RNe is thought to be powered by a thermonuclear runaway
(TNR) in the accreted layer formed on the white dwarf following accretion of 
mass from the companion (Starrfield et al 1985; Kato 1990; Kato 1991).

The outburst properties of the RNe with giant secondaries are quite homogeneous.
They are fast novae with a rate of decline of $\sim 0.3$~mag/day. The outburst
spectrum is characterized by broad emission lines ($V_{\rm{exp}}\sim 4000$~km
s$^{-1}$), which narrow with time, presence of intense coronal lines and
other high excitation lines. The 1985 outburst of RS Oph was one of the best
studied events, with the outburst being recorded from X-rays to radio
wavelengths (see Rosino 1987 for a review on the outbursts of RS Oph). 
The coronal lines, X-ray and the radio emission arise in a
region heated by the shock interaction of the fast moving dense nova ejecta
with the pre-outburst slow moving red giant stellar wind. Some of the parameters
of these systems are listed in Table 1.

\begin{table*}
\caption{Parameters of recurrent novae with giant secondaries}
\begin{center}
\begin{tabular}{llllclllll}\hline
Name & $m_{\rm{max}}$ & $m_{\rm{min}}$ & $t_3$ & $<t_{\rm{rec}}>$ & $E(B-V)$
& dist & Giant & $P$ & Ref\\
 & & & days & yrs & & kpc & & days& \\
\hline
T CrB & 2.0 & 10.2 & 6.8 & 80 & 0.15 & 1.3 & M3\,III & 227.67 & 1,2\\
RS Oph & 5.0 & 11.5 & 9.5 & 22 & 0.70 & 1.6 & M0/2\,III & 460 & 3,4\\
V3890 Sgr & 8.2 & 17.0: & 17.0 & 28 & 0.5 & 5.2 & M5\,III & & 5\\
V745 Sco & 9.6 & 19.0: & 14.9 & 52 & 1.1 & 4.6 & M6\,III & & 5\\
\hline\\
\multicolumn{10}{l}{References --  1, Selvelli et al (1992); 2, Belczy\'nski \& Miko{\l}ajewska (1998);}\\
\multicolumn{10}{l}{3, Cassatella et al (1985); 4, Dobrzycka \& Kenyon (1984);
5, Harrison et al (1993).}\\
\end{tabular}
\end{center}
\end{table*}

At quiescence, the optical spectrum is dominated by that of the red giant with
emission lines predominantly due to H\,{\sc i} and He\,{\sc i}. With the 
exception of T CrB, the other objects also have lines due to Fe\,{\sc ii} 
and Ca\,{\sc ii}. He\,{\sc ii} lines
are either extremely weak or absent. Van Winckel et al (1993) classify these
systems as belonging to the symbiotic type S3, i.e.\ stars with a slow, very
dense wind, leading to a deep central reversal of the H$\alpha$ emission line.

Only T CrB and RS Oph have been observed in the ultraviolet (UV) using the 
International Ultraviolet Explorer (IUE). The UV spectra 
of T CrB show a complex structure with emission lines and shell-like absorption 
features superposed over a relatively hot continuum (Selvelli et al 1992). The 
UV spectra of RS Oph show a flat continuum with a few weak emission lines like 
C\,{\sc iii} 3130, N\,{\sc iii}] 1750 and C\,{\sc iv} 1550 (Dobrzycka et al 1996a).
The UV continuum as well as the emission line fluxes are variable in both the 
objects, and are strongly correlated. Variability in the $UBV$/visual
magnitudes, and the optical emission lines 
has also been reported (e.g.\ Kenyon \& Garcia 1986; Iijima 1990; Iijima et al 
1994; Anupama \& Prabhu 1991; Anupama 1997; Dobrzycka et al 1996a). 

The binary nature of T CrB was first established by Sanford (1949), and later 
by Kraft (1958) who refined Sanford's period estimate to 227.6 days, derived a 
total mass of the system of 5~\msun, and a mass ratio of 1.4 with the giant
being the more massive component. Kenyon \& Garcia (1986) obtained new radial 
velocity data for the giant and combining with the data of Sanford and Kraft, 
obtained a new orbital solution and confirmed the previous estimates for the 
component masses. These results suggested the hot component has a mass exceeding 
the Chandrasekhar limit and hence must be a main sequence star, and posed
a serious problem for the thermonuclear runaway model for the nova outbursts 
requiring a massive white dwarf. 
Recently, Belczy\'nski \& Miko{\l}ajewska (1998) have 
reanalyzed the photometric and radial velocity data and arrived at new 
parameters for the binary components. Their analysis shows the mass ratio of 
T CrB $q\equiv M_{\rm{g}}/M_{\rm{h}}\approx 0.6$ with stellar
masses $M_{\rm{g}}\sim 0.7$~\msun\ for the red giant and $M_{\rm{h}}\sim
1.2$~\msun\ for the hot companion, compatible with a white dwarf mass. This 
is also in agreement with the UV characteristics of the hot component which 
are easily explained by the presence of a white dwarf acceptor (Selvelli et
al 1992).

Garcia (1986) found the radial velocity of the absorption features in the 
shell type profiles of the Fe\,{\sc ii} lines in RS Oph varied in a range from
$-30$~km s$^{-1}$ to $-50$~km s$^{-1}$ with a period of about 230 days.
Dobrzycka \& Kenyon (1994) recently reanalyzed the orbital period for
RS Oph and found the system has a spectroscopic orbit of period 460 days. 
Assuming the hot component to be a massive white dwarf, they estimated RS Oph
to be a low mass system, with low inclination, quite similar to T CrB.
The nature of the hot component in this system, however, still remains a bit 
of a mystery. Following the decline from the 1985 outburst, the temperature
and luminosity of the central ionizing source as inferred from the optical 
and X-ray spectra (Anupama \& Prabhu 1989; Mason et al 1987),
T$\sim 3.5\times 10^5$~K; L$\sim 10^{37}$ erg s$^{-1}$, were consistent with 
the remnant residual hydrogen burning on top of a white dwarf with
a thin atmosphere following the TNR outburst. The recent ROSAT detection of 
RS Oph at quiescence (Orio 1993) however, implies a soft X-ray luminosity 
$\sim 10^{31}-10^{32}$.  Recently, Dobrzycka et al (1996a) analyzed the 
quiescent optical and UV spectra and estimated the hot star has a luminosity 
of $L_h\sim 100-600$~\lsun\ and the spectrum mimics that of a B-type shell
star. The absence of He\,{\sc ii} lines in the spectra together with the 
presence of He\,{\sc i} lines led them to estimate the temperature of the 
source to be $\approx 5\times 10^4$~K. 
They also found that although the hot component luminosity is consistent
with the high accretion rate, $\sim 10^{-8}$~\msunyr, needed for
recurrent nova eruptions, the effective temperature
and luminosity place the hot component far from standard
massive white dwarf tracks in the HR diagram. 

In this work, the behaviour of these systems at quiescence is studied. The 
observed spectral features of the hot component are interpreted in terms of
an accreting white dwarf embedded in an envelope of wind from the giant 
secondary. 

\section{Observational Data and Results}

\begin{table}
\caption{Details of observations from VBO}
\begin{center}
\begin{tabular}{lllll}\hline
Date & JD$^*$ & Phase & $m_{5500}$ & $\lambda$ (\AA)\\
\hline
\multicolumn{5}{l}{T CrB$^1$}\\
1990, Mar 24.9 & 47975 & 0.47 & 10.2 & 4200--7000\\
1990, Mar 30.9 & 47981 & 0.50 & 10.4 & 4200--7000\\
1990, Mar 31.9 & 47982 & 0.50 & 10.3 & 4200--7000\\
1990, Apr 1.9 & 47983 & 0.51 & 10.4 & 6000--8800\\
1990, Apr 2.9 & 47984 & 0.51 & 10.2 & 6000--8800\\
1995, Mar 15.9 & 49792 & 0.45 & 10.0 & 4200--7000\\
1995, Apr 13.8 & 49821 & 0.58 & 10.3 & 4200--7000\\
1996, Apr 6.9 & 50180 & 0.16 & 9.7 & 4200--7000\\
1997, Apr 20.8 & 50559 & 0.82 & 10.0 & 3600--8000\\
1997, May 18.9 & 50587 & 0.94 & 9.8 & 3600--8000\\
1998, Mar 20.9 & 50893 & 0.29 & 9.9 & 3600--9000\\
\\
\multicolumn{5}{l}{RS Oph$^2$}\\
1990, Feb 21.0 & 47944 & 0.40 & 11.9 & 4300--7300\\
1990, Mar 31.7 & 47982 & 0.48 & 11.4 & 4500--8900\\
1991, Apr 15.8 & 48362 & 0.31 & 11.7 & 4400--9000\\
1992, Feb 14.0 & 48667 & 0.97 & 11.6 & 4100--7200\\ 
1992, Mar 13.9 & 48695 & 0.03 & 11.7 & 4100--7200\\
1993, Feb 15.0 & 49034 & 0.77 & 10.3 & 4200--7000\\
1993, Mar 28.0 & 49075 & 0.86 & 11.4 & 4000--7000\\
1995, Apr 13.9 & 49821 & 0.48 & 10.8 & 4500--7000\\
1996, Apr 6.9 & 50180 & 0.26 & 11.7 & 4200--7000\\
1997, Apr 20.9 & 50559 & 0.09 & 12.1 & 3600--8000\\
1997, May 18.8 & 50587 & 0.15 & 11.1 & 3600--8000\\
1998, Mar 19.4 & 50892 & 0.81 & 11.1 & 3600--9000\\
\\
\multicolumn{5}{l}{V3890 Sgr}\\
1997, Apr 20.9 & 50559 &    & 15.0 & 3600--8000\\
1998, Mar 21.0 & 50893 &    & 15.5 & 3600--8500\\
\\
\multicolumn{4}{l}{V745 Sco}\\
1998, Mar 20.9 & 50893 &    & 17.7 & 3600--8500\\
\hline\\
\multicolumn{4}{l}{*: JD\,2400000+}\\
\multicolumn{4}{l}{1: $T_0=JD\,2431931.05+227.67\,E$}\\
\multicolumn{4}{l}{2: $T_0=JD\,2444999.9+460\,E$}\\
\end{tabular}
\end{center}
\end{table}

All published photometric data of these objects at quiescence, as well as the 
visual magnitude observations from AAVSO and VSNET have been used in this study.
Spectroscopic data are based on those published in the literature, as well as
CCD spectra obtained from the Vainu Bappu Observatory (VBO) using both the 
1.02m and the 2.3m telescopes. The VBO spectra were
obtained during 1990--1998 at 10--12~\AA\ resolution. The details of observations
are given in Table 2. All spectra were reduced in the 
standard method, and brought to flux scale using spectrophotometric standards. 

\begin{table*}
\caption{T CrB: Emission line fluxes in $10^{-12}$ erg
cm$^{-2}$ s$^{-1}$}
\begin{center}
\begin{tabular}{lrrrrrrrrrrr}\hline
&\multicolumn{11}{c}{JD 2400000+}\\
$\lambda$~\AA&47975&47981&47982&47983&47984&49792&49821&50180&50559&50587&50893\\
\hline
H$\delta$ 4101&      &      &      & & &      &      &      & 0.74 & 1.48 &\\
H$\gamma$ 4340&      &      &      & & &      &      & 1.48 & 0.94 & 1.52 &\\
He\,{\sc ii} 4686&   &      &      & & &      &      &      & 0.35 & 0.43 &\\
H$\beta$ 4861& 0.68 & 1.28 & 1.06 & & & 0.92 & 0.14 & 3.08 & 2.34 & 3.75 & 0.59\\
He\,{\sc i} 5876&   &      &      & & &      &      & 1.02 & 0.86 & 1.22 &\\
H$\alpha$ 6563& 2.36 & 5.20 & 5.00 & 5.91 & 6.46 & 5.07 & 1.21 & 13.78 & 6.36 & 8.60 & 3.71\\
He\,{\sc i} 6678&   &      &      & & &      &      & 1.24 & 0.58 & 1.19 &\\
He\,{\sc i} 7065&   &      &      & & &      &      &      & 0.46 &      &\\
\hline\\
\end{tabular}
\end{center}
\end{table*}

\begin{table*}
\caption{RS Oph: Emission line fluxes in $10^{-12}$ erg cm$^{-2}$ s$^{-1}$}
\begin{center}
\begin{tabular}{lrrrrrrrrrrrr}\hline

 & \multicolumn{12}{c}{JD 2400000+}\\
$\lambda$~\AA &47944&47982&48362&48667&48695&49034&49075&49821&50180&50559&50587&50892\\
\hline
H$\delta$ 4101&   &    &    &    &    &    &0.19& & & & &0.03\\
H$\gamma$ 4340&   &    &    &0.15&0.33&0.48&0.37& &0.14&0.12&0.32&0.04\\
H$\beta$ 4861&0.76&1.18&0.87&0.36&0.81&2.75&1.09&1.73&0.86&0.81&1.50&0.46\\
Fe\,{\sc ii} 4924&0.19&0.23&0.13&0.06&0.10&0.27&0.13&0.16&0.14&0.10&0.13&0.06\\
Fe\,{\sc ii} 5018&0.13&0.30&0.14&0.04&0.13&0.45&0.13&0.24&0.22&0.27&0.23&0.09\\
Fe\,{\sc ii} 5169&0.16&0.11&0.21&0.12&0.16&0.63&0.32&0.16&0.15&0.14&0.45&0.09\\
Fe\,{\sc ii} 5235&0.12&0.15&0.08& &0.06&0.15&0.07& &0.13&0.05&0.14&0.04\\
Fe\,{\sc ii} 5276&0.04&0.23&0.13&0.05&0.14&0.74&0.30&0.42& &0.18&0.29&\\
Fe\,{\sc ii} 5317&0.28&0.34&0.20&0.10&0.08&0.86&0.27&0.14&0.12&0.15&0.33&0.08\\
Fe\,{\sc ii} 5363&0.22&0.28&0.10& &0.11&0.45&0.07&0.15&0.07&0.03&0.16&0.07\\
Fe\,{\sc ii} 5535&0.05&0.13& & &0.13&0.19&0.09&0.16&0.06&0.07&0.10&0.12\\
He\,{\sc i} 5876&0.33&0.76&0.56&0.24&0.58&1.30&0.54&0.81&0.54&0.36&0.81&0.42\\
Fe\,{\sc ii} 5991&0.08&0.12&0.07& &0.09&0.14& & &0.08&0.06&0.15&\\
H$\alpha$ 6563&6.93&11.70&9.31&4.19&1.12&25.88&10.57&17.10&11.79&7.43&14.89&15.19\\
He\,{\sc i} 6678&0.20&0.39&0.30&0.10&0.50&0.48&0.15&0.25&0.33&0.18&0.21&0.27\\
He\,{\sc i} 7065&0.33&0.74&0.43&0.13&0.36& & & & &0.22&0.26&0.15\\
O\,{\sc i} 7774& & &0.09& & & & & & &0.04& &0.13\\ 
O\,{\sc i} 8446& &1.01&0.93& & & & & & & & &1.50\\
Ca\,{\sc ii} 8498& &0.88&0.87& & & & & & & & &1.17\\
Ca\,{\sc ii} 8542& &0.45&0.64& & & & & & & & &0.97\\
Ca\,{\sc ii} 8662& &0.51&0.63& & & & & & & & &0.75\\
\hline\\
\end{tabular}
\end{center}
\end{table*}

In Figures 1--4 we
show, respectively, the spectra of T CrB, RS Oph, V3890 Sgr and V745 Sco.
The fluxes of selected emission lines are listed in Tables 3--6. The errors
in the fluxes are $\sim 7-10$\% for the bright lines and $\sim 15-20$\% for the
fainter features. The errors in some of the faint features in V745 Sco could be 
around 25\%. The flux calibration is inaccurate beyond 8400~\AA\ in the spectra
of V3890 Sgr and V745 Sco obtained in 1998.

\begin{figure}
\epsfysize 8cm \epsfbox{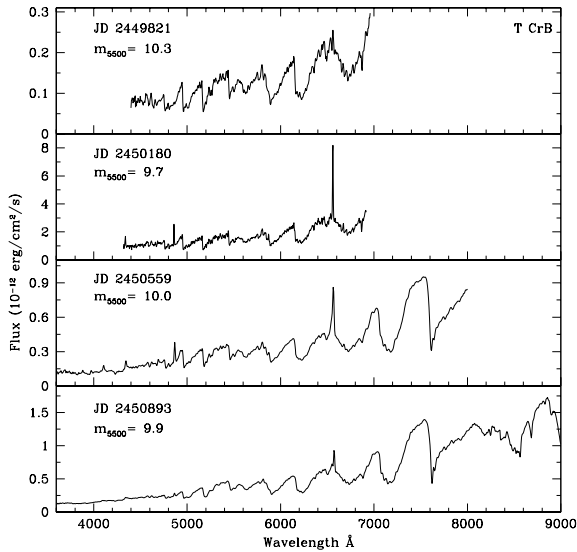}
\caption{Sample spectra of T CrB based on observations from VBO.}
\end{figure}

\begin{figure}
\epsfysize 8cm \epsfbox{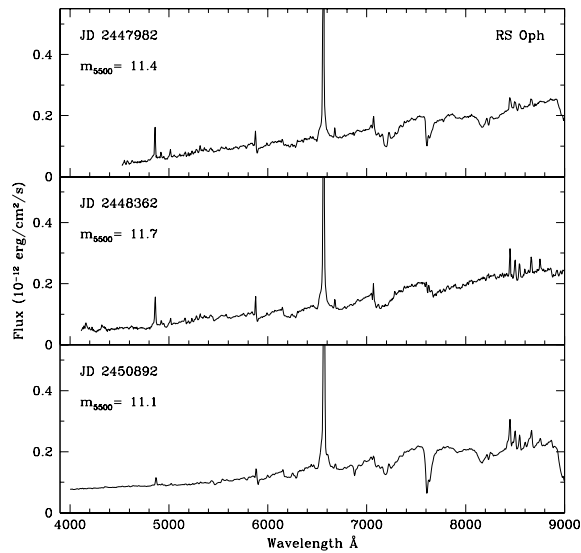}
\caption{Sample spectra of RS Oph obtained from VBO. The H$\alpha$ line is 
truncated.}
\end{figure}

\begin{figure}
\epsfysize 7cm \epsfxsize 8cm \epsfbox{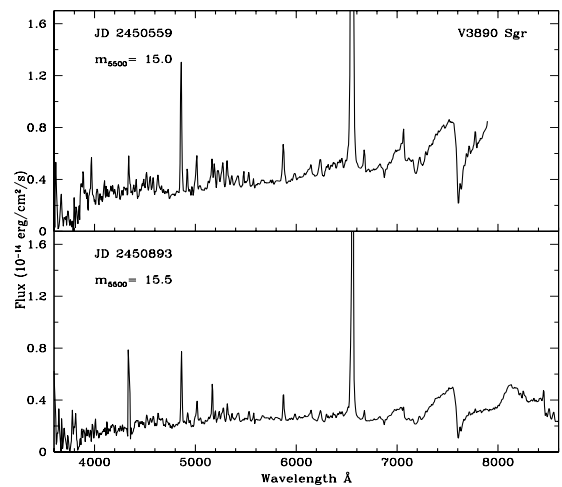}
\caption{Spectra of V3890 Sgr obtained from VBO. The H$\alpha$ line is
truncated.}
\end{figure}

\begin{figure}
\epsfysize 5cm \epsfxsize 8cm \epsfbox{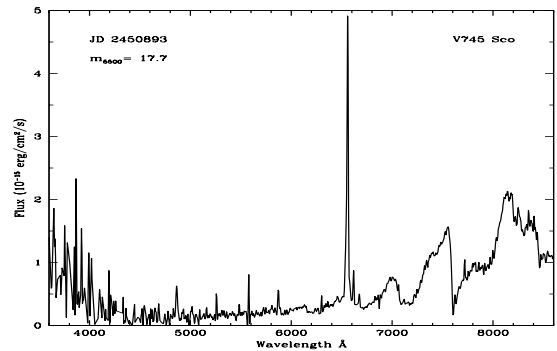}
\caption{Spectrum of V745 Sco.}
\end{figure}

\section{Analysis}
\begin{table}
\caption{V3890 Sgr: Emission line fluxes in $10^{-13}$ erg cm$^{-2}$ s$^{-1}$}
\begin{center}
\begin{tabular}{lrr}\hline
&\multicolumn{2}{c}{JD 2400000+}\\
$\lambda$~\AA &50559&50893\\
\hline
H$\gamma$ 4340&0.37&0.26\\
He\,{\sc ii}4685&0.01&\\
H$\beta$ 4861&1.48&0.88\\
Fe\,{\sc ii} 4924&0.24&0.14\\
Fe\,{\sc ii} 5018&0.56&\\
He\,{\sc i} 5876&0.60&0.35\\
H$\alpha$ 6563&12.21&5.87\\
He\,{\sc i} 6678&0.27&0.12\\
He\,{\sc i} 7065&0.25&0.08\\
O\,{\sc i} 7774&0.22&0.03\\
O\,{\sc i} 8446& &1.45\\
Ca\,{\sc ii} 8498& &0.12\\
Ca\,{\sc ii} 8542& &0.11\\
\hline\\
\end{tabular}
\end{center}
\end{table}

\begin{table}
\caption{V745 Sco: Emission line fluxes in $10^{-14}$ erg cm$^{-2}$ s$^{-1}$}
\begin{center}
\begin{tabular}{lr}\hline
\multicolumn{2}{c}{JD 2450893}\\
$\lambda$~\AA& Flux\\
\hline
H$\beta$ 4861&1.89\\
Fe\,{\sc ii} 4924&0.27\\
Fe\,{\sc ii} 5018&0.32\\
Fe\,{\sc ii} 5169&0.35\\
Fe\,{\sc ii} 5317&0.20\\
Fe\,{\sc ii} 5363&0.19\\
He\,{\sc i} 5876&1.15\\
H$\alpha$ 6563&12.92\\
He\,{\sc i} 6678&0.46\\
He\,{\sc i} 7065&0.35\\
\hline\\
\end{tabular}
\end{center}
\end{table}

\subsection{Variability}
\subsubsection{T Coronae Borealis}

The $UBV$ and $J$ light curves of T CrB (Belczy\'nski \& Miko{\l}ajewska 1998 and
references therein) show sinusoidal variations with half the orbital period 
caused by the orbital motion of the tidally distorted red giant. This effect is
more pronounced in the $V$ and $J$ bands, while in the $B$ and $U$ bands it is
superposed upon secular changes. The $U$ light curve is dominated by the
secular changes, as well as some erratic variations, which can be attributed
to the hot component.

The optical emission line fluxes vary considerably during the period covered 
by our observations (Table 3).
Similar variability has been previously observed 
(e.g. Andrillat \& Houziaux 1982; 
Kenyon \& Garcia 1986; Iijima 1990; Anupama \& Prabhu 1991; Anupama 
1997). 
Anupama (1997) also reports long-term periodicities in the optical emission
line variation. 
T CrB was in a high state in 1987 with the emission lines being
very bright, followed by a low state in the early 1990's, when only the Balmer
H$\alpha$ and H$\beta$ lines were weakly present. The emission line strengths
increased once again in 1996, accompanied by significant brightening
of the optical continuum.

Figure 5 (top panel) shows our H$\alpha$ and H$\beta$ line fluxes 
together with those published plotted against the $V$ magnitude.
The correlation between the line strengths and the $V$ magnitude is evident. 
Iijima (1990) found a similar behaviour of the H$\beta$ flux with 
the $B$ magnitude. 
The quiescent $V$ light curve of T CrB is dominated by the ellipsoidal changes
of the cool giant. 
To estimate the $V$ magnitude of the hot component, $m_{\rm hot}$, 
we have subtracted the contribution of the M giant to the observed $V$ 
magnitudes of T CrB using the synthetic ellipsoidal light curve for the 
M giant computed by Belczy\'nski \& Miko{\l}ajewska (1998). 
The strong correlation of the emission line fluxes with
so estimated $m_{\rm hot}$ (bottom panel of Figure 5) indicates that
the hot component activity is responsible for the optical emission line changes.

\begin{figure}
\epsfysize 8cm \epsfbox{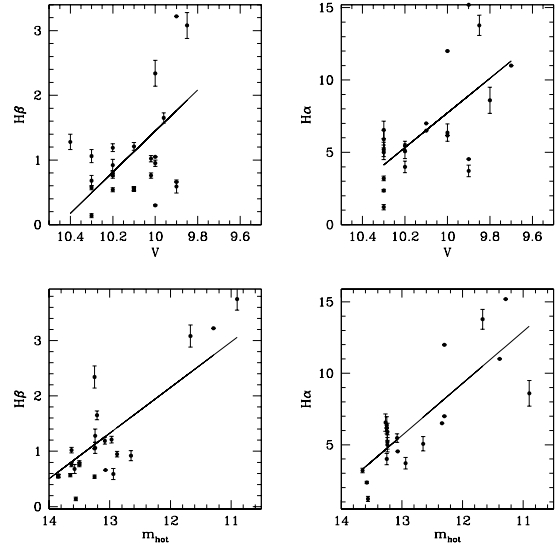}
\caption{Correlation of the H$\beta$ and H$\alpha$ emission line fluxes
in T CrB with the $V$ magnitude (top panel), and the $V$ magnitude of the hot
component $m_{\rm{hot}}$ (bottom panel). Fluxes are in $10^{-12}$ erg cm$^{-2}$
s$^{-1}$.}
\end{figure}

The enhancement in the line fluxes in the optical are also correlated with the
activity in the UV. For example, the fluxes in the UV and the optical emission
lines were low in 1981 (Williams 1983, Blair et al 1983, Selvelli et al 1992),
while the high state observed in the optical in 1987 coincides 
with the enhancement seen in the June 1987 UV spectrum. During the period
1982--1990, Luthardt (1992) found the variations in the mean $U$ brightness
by more than 2.5~mag, while the variations in the $B$ and $V$ band were,
respectively, about 1~mag and 0.3~mag. The variation in the $U$ band is
strongly correlated with the variation observed in the UV flux 
during the same period. It is thus quite evident that the optical 
enhancements seen in T CrB are associated with the hot component. We will
discuss later as to the possible causes of the activity.

\subsubsection{RS Ophiuchi}

The visual magnitude of RS Ophiuchi is found to vary between 11--12 magnitude,
with occasional rises to $\sim 10$ magnitude. The spectra presented in Figure 2 
and the fluxes listed in Table 4, indicate that the optical emission lines are 
also variable. Variability in the UV flux has also been reported (Dobrzycka et al
1996a, Shore et al 1996). Figure 6 presents our H$\beta$ and H$\alpha$ 
flux estimates together with published data plotted against 
the continuum magnitude at 5500 $\rm \AA$, $m_{5500}$, 
dependence of the Balmer line ratios on $m_{5500}$, 
and the $F_{\rm{UV}}$ (from Shore et al 1996) vs. $m_{\rm{vis}}$.
The figure shows that the emission line flux 
variability is strongly correlated with the activity of the hot component as in
the case of T CrB. Although the data available in the UV are few, 
there is also an indication of a correlation between the optical and UV flux.

\begin{figure}
\epsfysize 8cm \epsfbox{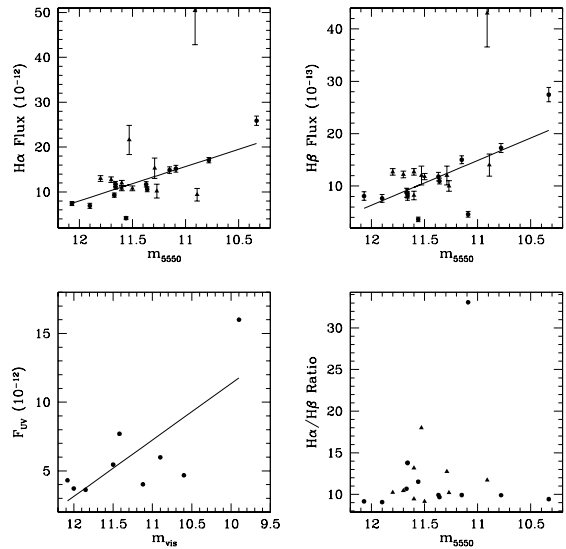}
\caption{Top: Correlation of the H$\beta$ and H$\alpha$ emission line
fluxes in RS Oph with the monochromatic magnitude at 5500~\AA.
Bottom: Correlation of the total UV flux with visual magnitudes (left) and the
observed Balmer line ratios as a function of $m_{5500}$ (right). All fluxes are
in erg cm$^{-2}$ s$^{-1}$. Filled circles represent VBO data and filled triangles
represent published data.}
\end{figure}

Broad emission components in the emission lines of H\,{\sc i} and He\,{\sc i} 
were detected by Iijima et al (1994) during 1991 and 1992. These broad components
are present in the H$\beta$ and H$\alpha$ lines in some of our spectra 
(see Figure 2) at velocities of the order of $\sim 1000$~km s$^{-1}$.  The 
wavelength, profile and intensity of these components are found to be strongly 
variable with a time scale of days (Iijima et al 1994). 
Iijima et al also detect He\,{\sc ii} 4686 at an intensity of about 8\% of 
H$\beta$ on 1991 Oct 2 and suggest the possibility of a relation between the
presence of this line and the intensity of the broad components.  They also 
suggest that the broad components arise in a region close to the surface of the 
hot component, a white dwarf. 

\subsubsection{V3890 Sagittari and V745 Scorpii}

Similar to RS Oph and T CrB, V3890 Sgr also shows activity at quiescence, as 
depicted by the recent rise to $\sim 14.6$~mag in May 1997. 
The spectrum of 1997 (Figure 3) was obtained 
when V3890 Sgr was in a high state at a 
visual magnitude of 15 mag. The emission lines are stronger in 1997 as compared 
to 1998 (Figure 3; also Table 5). Comparing the spectra presented here with
that of Williams (1983) obtained in 1981, it appears that this system was in a
high state also in 1981. The emission lines were very strong and the ionization 
level was higher than that indicated by the 1997 and 1998 spectra.
For example, 
He\,{\sc ii} 4686~\AA\ was present with a strength $\sim 28$\% that of
H$\beta$ in 1981, while it is barely detected in 1997 and absent in 1998.
It is interesting to note that the He\,{\sc ii} 4686 line was quite strong
in the spectra presented by Williams et al (1994) obtained 
1--2 years after the 1990 outburst.

Figure 4 shows the spectrum of V745 Sco obtained in March 1998. Although the 
blue region of the spectrum is very noisy, the lines due to H$\beta$, 
Fe\,{\sc ii} and He\,{\sc i} 5876 can be easily identified. We are
unable to say anything about activity in this nova owing to inadequate data.
However, we expect its behaviour to be similar to that of the other members in
the group.

\subsection{Flickering activity}

Rapid photometric variations with amplitudes of 0.01--1.0 magnitude over
timescales of minutes have been observed in both T CrB and RS Oph. The
amplitude of flickering in T CrB is maximum in the $U$ band with $\Delta U
\sim 0.1-0.5$ magnitude on timescales of minutes (Walker 1957; Ianna 1964;
Lawerence et al 1967; Bianchini \& Middleditch 1976; Walker 1977; Bruch 1980;
Oskanian 1983). The amplitude is smaller in $B$, $\Delta B\sim 0.1-0.3$ mag,
and generally undetectable in the $VRI$ bands (Raikova \& Antov 1986; Lines et al
1988; Bruch 1992; Miko{\l}ajewski et al 1997). On some occasions, there is no
flickering detected at all wavelengths (Bianchini \& Middleditch 1976; 
Oskanian 1983; Dobrzycka et al 1996b; Miko{\l}ajewski et al 1997). The absence
of flickering in the longer wavelength bands dominated by flux from the cool
component indicates these rapid fluctuations are associated with the hot
component. 

We have looked for correlation of flickering with the activity
of the hot component. Flickering is generally absent when the system is 
in a low state and usually present when the system is in a high state. 
In particular, T CrB was is in a relatively high state during most
of 1981-86 period (e.g. Fig. 1 of Belczy{\'n}ski \& Miko{\l}ajewska 1998),
when various observers detected flickering in the $U$, $B$ and occasionally 
even in the $V$ light (Oskanian 1983; Lines et al 1988; Bruch 1992; Raikova \& Antov
1986) with amplitudes up to 0.6 mag in $U$. 
The flickering activity however disappeared in June 1982,
following a significant decline in $UBV$ magnitudes ($\Delta U \sim 0.7$,
$\Delta B \sim 0.6$ and $\Delta V \sim 0.5$). 
Similarly, the absence of flickering in June 1993 (Dobrzycka et al
1996b) coincided with a general drop in the $UBV$ as well as IUE fluxes
after c. JD 2\,447\,000 (Luthardt 1992; Selvelli et al. 1992;
Belczy{\'n}ski \& Miko{\l}ajewska 1998). This low state prevailed
until spring 1996; flickering reappeared in April 1996, when T CrB brightened
significantly, and was generally present when the star remained
blue ($B-V \la 1.2$, $U-B \la 0.3$; Fig. 1 of Miko{\l}ajewski et al 1997).
The presence or absence of flickering does not show any correlation 
with the orbital phase. It is however, evident that flickering is correlated
with the activity of the hot component.

The observations of flickering activity in RS Oph are few.
However, all observations show a consistent flickering activity with amplitudes
of $\Delta U \sim 0.2-0.3$ and $\Delta B \sim 0.3$ (Walker 1957; Walker 1977;
Bruch 1980, 1992; Dobrzycka et al 1996b) over timescales of minutes.
We must however note that all observations were made when the star
was relatively bright ($V \la 11.45$) and blue ($B-V \la 1.1$).

The flickering activity detected in both T CrB and RS Oph is very similar 
to that observed in dwarf novae and other cataclysmic variables containing
accreting white dwarfs, and provides very strong evidence for the presence
of accreting white dwarfs in these recurrent nova systems also.

\subsection{The cool component}

Kenyon \& Fernandez-Castro (1987) showed that
the red TiO bands at 6180~\AA\ and 7100~\AA\ together
the VO 7865~\AA\ band and the Na\,{\sc i} 
infrared doublet at 8190~\AA\ provide good diagnostics for K-M stars.
Thus, to estimate the spectral type and luminosity class of the giant secondary
in our target systems, we have used the [TiO]$_{1,2}$, 
[VO] and [Na\,{\sc i}] indices 
as defined by Kenyon \& Fernandez-Castro. Our results are given in Table 7.

\begin{table}
\caption{Absorption indices and secondary spectral type}
\begin{center}
\begin{tabular}{lllrrl}\hline
JD$^*$ & [TiO]$_1$ & [TiO]$_2$ & [VO] & [Na\,{\sc i}] & Sp. type\\
\hline
\multicolumn{6}{l}{T CrB}\\
50180 & 0.59 & & & & M4\,III\\
50559 & 0.55 & 0.73 & & & M3/4\,III\\
50587 & 0.57 & 0.81 & & & M4\,III\\
50893 & 0.49 & 0.64 & 0.16 & $-0.03$ & M3\,III\\
\\
\multicolumn{6}{l}{RS Oph}\\
47944 & 0.15 & 0.12 & & & K5.3/K5.6\\
47982 & 0.13 & 0.17 & $-0.11$ & 0.82 &K5-M0\,III\\
48362 & 0.22 & 0.31 & 0.02 & 0.98 & M0-M1\,III\\
48665 & 0.12 &      &      &      & K5.3\\
48695 & 0.04 &      &      &      & K4.7\\
49034 & 0.13 &      &      &      & K5.4\\
49075 & 0.16 &      &      &      & K5.7\\
49821 & 0.14 & 0.15 &      &      & K5.6\\
50180 & 0.16 &      &      &      & K5.7\\
50559 & 0.09 & 0.13 &      &      & K5/K5.4\\
50587 & 0.16 & 0.29 &      &      & K5.7/M1\\
50892 & 0.15 & 0.18 & $-0.18$ & 0.18 & K5.6-M0\,III\\
\\
\multicolumn{6}{l}{V3890 Sgr}\\
50559 & 0.13 & 0.20 & & & K5.5/M0\\
50893 & 0.15 & 0.36 & 0.27 & $-0.02$ & K5.6/M1.5/M5\,III\\
\\
\multicolumn{6}{l}{V745 Sco}\\
50893 & 0.41 & 0.90 & 0.41 & $-0.31$ & M2/M5/M6\,III\\
\hline\\
\multicolumn{6}{l}{*: JD\,2400000+}
\end{tabular}
\end{center}
\end{table}

\noindent
{\it T CrB:}\newline
Our spectra indicate a spectral type of M3-4\,III for the M giant in T CrB
consistent with previous estimates (Kenyon \& Fernandez-Castro 1987, Webbink
et al 1987) based on optical spectra. The $K$ band spectrum and the infrared
colours are also consistent with this spectral classification (Harrison et al
1993). The M giant does not show any pulsations. 

\noindent
{\it RS Oph:}\newline
Previous spectral classifications for the cool component in RS Oph range from
G5--M2\,III. Our spectra indicate a spectral type M$0\pm 1$, with
the luminosity classification consistent with that of a giant. The infrared
$K$ band spectra are also consistent with an M0\,III star (Evans et al 1988,
Scott et al 1994).

Our results (Table 7) show that the secondary spectral type varies
between K4--M1. Moreover, the [TiO]$_1$ index indicates an earlier spectral type
compared to the [TiO]$_2$ index, indicating a presence of an additional blue
continuum at shorter wavelengths. This blue continuum must be produced
by the hot component, since the nebular continuum as predicted
by the observed H$\beta$ flux is too faint to produce any measurable effect. 
The dilution of the TiO indices 
due to the hot component contribution is clearly evident in the data presented 
by Dobrzycka et al (1996a). 
In particular, the spectrum obtained 120 days after the 1985 outburst,
when the hot component contribution to the optical/red continuum flux
was negligible, 
indicates an M2\,III spectral type, while spectra obtained during other periods
indicate spectral type ranging between K5--M0\,III, similar to our observations.

\noindent
{\it V3890 Sgr:}\newline
Based on the TiO and VO features between 7000~\AA\ and 8000~\AA\ in the
immediate post-1990 outburst, Williams et al (1991) classify the secondary
spectral type as M8\,III. Comparing our spectra obtained in 1997 and 1998
with those presented by Williams et al, we find the strengths of the TiO
bands have decreased. 

Similar to RS Oph, the spectral type indicated by the longer wavelength indices
are later (Table 7). 
The [TiO]$_1$ index implies a K5 type, while the [VO] and
the [Na\,{\sc i}] indices indicate a spectral type M5\,III. The infrared
colours of V3890 Sgr based on photometry in 1991 (Harrison et al 1993) also
indicate a spectral type M5\,III.

\noindent
{\it V745 Sco:}\newline
The secondary in V745 Sco has been classified as M6--8\,III (Sekiguchi et al
1990; Duerbeck \& Seitter 1989; Williams et al 1991). The infrared
spectra and colours indicate a spectral type M4\,III (Harrison et al 1993).
The absorption indices derived in this study indicate 
significant contribution from the hot component as
in the case of both RS Oph and V3890 Sgr. In particular, the [TiO]$_2$
and [VO] indices indicate a spectral type M4--5\,III, while the [Na\,{\sc i}]
index is consistent with an M6 giant. 

\subsection{Spectral decomposition}

\begin{figure}
\epsfysize 8cm \epsfbox{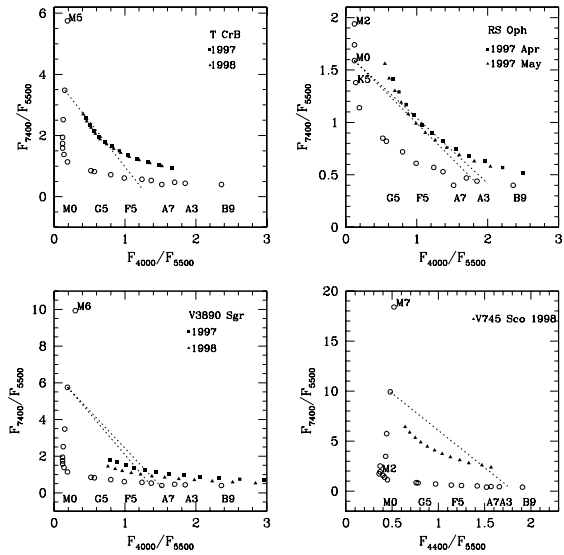}
\caption{The continuum flux ratio diagram. Open circles represent normal
stars. The ratios for T CrB (top left), RS Oph (top right), V3890 Sgr (bottom 
left) and V745 Sco (bottom right) are plotted for different $E(B-V)$ ranging
from 0.1 to 1.2, from top to bottom, in steps of 0.1. Also shown are the lines
connecting the giant and the hot component spectral type for the adopted 
$E(B-V)$.}
\end{figure}

In the previous section we have found that the optical spectrum
includes flux from the cool giant as well as significant (especially
in the case of RS Oph, V3890 Sgr and V745 Sco) contribution from the hot
component, while the nebular continuum is generally negligible.
Our next goal is to determine the temperature and luminosity of the hot
component from our optical data using simple spectral decomposition techniques.

Figure 7 shows the observed flux ratio
$F_{7400}/F_{5500}$ plotted against the ratio $F_{4000}/F_{5500}$ for all the 
four objects. The observed flux ratios are reddening corrected for a range of 
$E(B-V)=0.1-1.2$. Also plotted in the same figure are the standard flux ratios 
for normal stars. 
In the case of V745 Sco and RS Oph spectra obtained in 1998, 
we use the ratio $F_{4400}/F_{5500}$
instead of $F_{4000}/F_{5500}$ as the region around 4000~\AA\ in these data is
noisy and the fluxes inaccurate. 
Since the flux ratios for a composite system lie on a straight line connecting
flux ratios for the individual components, the relative lengths of the line
segments give the fractional contribution of each component at the common
wavelength, 5500~\AA\ (Wade 1982; Dobrzycka et al 1996a).
Table 8 lists results from our flux-ratio diagrams,
the fractional contribution 
($f_{\rm h}$), spectral type (SP$_{\rm h}$) 
and magnitude ($V_{\rm h}^0$) of the 
hot component, and the magnitude of the giant ($V_{\rm g}^0$),
respectively, assuming the giant 
component of each system has the spectral type (SP$_{\rm g}$)
as estimated from the absorption indices (sec. 3.3). 
The adopted reddening value is also listed in the table. 

In the case of T CrB, 
we find that although the 7400/5500 ratio (Figure 7, $E(B-V)=0.15$) is 
consistent with that of an M3 giant, 
the 4000/5500 ratio is higher than a normal 
M3 giant. This indicates significant contribution from the hot component at 
lower wavelengths, consistent with the brightening observed in 1996--1997. 

The results obtained for RS Oph (Table 8) suggest that the on JD\,2450559
the visual brightness of both
the giant and the hot components was lower than usually.
In fact, the [TiO] indices (Table 7) indicate a somewhat earlier
spectral type for the giant than usually, possibly due to enhanced
veiling by the hot component continuum, while the emission line fluxes
(Table 4) are consistent with the
rather low luminosity of the hot component. 
It seems that the giant was simply in a low state on that day. 
Low states of the M giant are also observed in the highly variable 
symbiotic system CH Cyg, whose giant however is much cooler ($\ga$ M6) 
than the giant in RS Oph.

\begin{table}
\caption{Results of spectral decomposition}
\begin{center}
\begin{tabular}{llllllc}\hline
JD$^*$ & SP$_{\rm g}$ & SP$_{\rm h}$ & $f_{\rm h}$ & $V_{\rm h}^0$ & 
$V_{\rm g}^0$ & $E(B-V)$\\
\hline
\multicolumn{7}{l}{T CrB}\\
Average$^1$ & M4 & F0/F3 & 0.20- & 10.9 & 9.7 & 0.15\\
            &    &       & 0.25 &      &      &     \\
\\
\multicolumn{7}{l}{RS Oph}\\
50559 & M0/2 & A2/A4 & 0.60 & 10.5 & 10.9 & 0.7\\
50587 & M0/2 & A2/A4 & 0.57 & 9.5 & 9.8 & 0.7\\
50892 & M0/2 & A0/B9 & 0.55 & 9.6 & 9.8 & 0.7\\
\\
\multicolumn{7}{l}{V3890 Sgr}\\
50559 & M5 & A7 & 0.84 & 13.6 & 15.4 & 0.5\\
50893 & M5 & F0 & 0.88 & 14.1 & 16.2 & 0.5\\
\\
\multicolumn{7}{l}{V745 Sco}\\
50893 & M6 & A1 & 0.75 & 14.6 & 15.8 & 1.1\\
\hline\\
\multicolumn{7}{l}{*: JD\,2400000+}\\
\multicolumn{7}{l}{1: Average of JD\,2450559 and JD\,2450587}\\
\end{tabular}
\end{center}
\end{table}

The hot component mimics the
spectrum of an F0--F3 star in T CrB, a B9--A4 star in RS Oph 
(see also Dobrzycka et al 1996a), A7--F0 in V3890 Sgr and A1 in V745 Sco. 
The optical spectral types combined with $V_{\rm h}$ magnitudes
correspond to the hot component luminosity,
$L_h\sim 50\,(d/1.3\, \rm kpc)^2$~\lsun\ for T CrB,
$L_h\sim 190\,(d/1.5\, \rm kpc)^2$~\lsun\ for RS Oph, 
$L_h\sim 50-70\,(d/5\, \rm kpc)^2$~\lsun\ for V3890 Sgr, and
$L_h\sim 25\,(d/5\, \rm kpc)^2$~\lsun\ for V745 Sco, respectively. 

A check on the estimated temperature and luminosity of the hot component can be
obtained by an estimate of the total luminosity of the source below the Lyman
limit approximated as the sum of the H\,{\sc i}, He\,{\sc i} and He\,{\sc ii} 
Lyman continua inferred from the luminosities in H$\beta$, He\,{\sc i} 5876
and He\,{\sc ii} 4686 emission lines (Kenyon et al 1991):
$$L_{{\rm{EUV}}}\sim 50\,L({\rm{H}}\beta)+105\,L({\rm{He\,{\sc i}}}5876)
+110\,L({\rm{He\,{\sc ii}}}4686).$$
The observed line fluxes, corrected for the reddening values listed in Table 1, imply 
$L_{\rm{EUV}}\sim 35\,(d/1.3\, \rm kpc)^2$~\lsun\ in T CrB, 
$L_{\rm{EUV}}\sim 100-600\,(d/1.5\, \rm kpc)^2$~\lsun\ in RS Oph, 
$L_{\rm{EUV}}\sim 30\,(d/5\, \rm kpc)^2$~\lsun\ in V3890 Sgr 
and $L_{\rm{EUV}}\sim 40\,(d/5\, \rm kpc)^2$~\lsun\ in V745 Sco. 
In all cases, the EUV luminosities inferred from the emission line fluxes are
much higher, an order of magnitude or so, than expected for normal stars
with the spectral types derived from our spectral decomposition,
and is comparable to the UV/optical flux from the hot component.
Our results are similar to that obtained for RS Oph 
by Dobrzycka et al (1996a).

\subsection{OI 8446~\AA\ line}

The O\,{\sc i} 8446~\AA\ line is observed in classical novae during the 
diffusion-enhanced and Orion phases. The flux of the O\,{\sc i} 8446 line 
compared to that of other O\,{\sc i} lines such as 6300~\AA\ and 7774~\AA\ 
indicates the line is enhanced by Ly$\beta$ fluorescence. Ly$\beta$ enhanced 
O\,{\sc i} 8446 line has been detected in
the outburst spectra of RS Oph, V3890 Sgr and V745 Sco. The line strength 
reaches a maximum during the coronal line phases and begins to decline
thereafter as the density decreases (Anupama \& Prabhu 1989; Williams et al
1991). The line is absent during the nebular stages of the outburst. 
In RS Oph, the 8446~\AA\ line appears once again in the spectrum during
quiescence as seen in the 
data presented here as well as the spectra presented in Anupama \& Prabhu 
(1990). On the other hand, the O\,{\sc i} 7774~\AA\ line is very weak or absent.
O\,{\sc i} 8446 line is also present in V3890 Sgr and could be weakly present in 
V745 Sco. The T CrB spectra presented here do not show this line. However,
O\,{\sc i} 1304~\AA\ the third cascade line is seen in the IUE spectra (Selvelli 
et al 1992), especially when the object has been in a high state. Unfortunately,
our data during the recent high state of T CrB in 1996/1997 does not cover the
8446~\AA\ region.

Kastner \& Bhatia (1995) have recently calculated the fluorescent line
intensities expected by Ly$\beta$ photoexcitation of the oxygen spectrum. 
The Fe\,{\sc ii} lines and the Ca\,{\sc ii} near infrared triplet lines in RS Oph
imply an electron density of $\ga 10^{11}$~cm$^{-3}$ (see section 4.1). Assuming 
the O\,{\sc i} 
line arises in the same region, and comparing the observed ratio O\,{\sc i} 
$8446/7774\sim 10$ with table 5 and figure 8 of Kastner \& Bhatia, we estimate 
the photexcitation rate to be in the range $R_p\sim 10^{-2}-1$~s$^{-1}$. This 
rate corresponds to a mean Ly$\beta$ intensity of 
$J_{\rm{Ly}\beta}\ga 1240$~erg cm$^{-2}$ s$^{-1}$ sr$^{-1}$. Although the 
8446~\AA\ flux is inaccurate in V3890 Sgr, we can estimate a lower limit to 
the ratio as $\sim 3$, corresponding to $R_p\ga 10^{-2}$, similar to RS Oph. 
This implies the existence of similar conditions in the line emitting regions 
as well as the ionizing source in both RS Oph and V3890 Sgr. The value of the 
Ly$\beta$ intensity obtained is a lower limit as the theoretical calculations 
are for the optically thin case while the line emitting region is optically 
thick as seen from the Balmer line ratios in these objects. 

\section{Discussion}
\subsection{The hot component}

Although current thoughts favor a thermonuclear runaway on a massive
white dwarf as the energy source for recurrent novae,
there is a considerable debate in the literature
about the nature of the hot component in T CrB.
Basing on a rather controversial estimate of the hot component mass
above the Chandrasekhar limit (Kraft 1958; Kenyon \& Garcia 1986),
Webbink et al (1987) and 
Canizzo \& Kenyon (1992) interpreted the nova-like outbursts of T CrB in
terms of transient phenomena in a non-stationary accretion disc around a
main sequence star.
Unfortunately, as demonstrated by Selvelli et al (1992)
most observational data for T CrB are against the accretion model.
In particular, they showed that the quiescent 
UV characteristics of T CrB 
provide direct observational evidence for the presence of a white dwarf 
accreting at rates $\dot M_{\rm{acc}}\approx 10^{-8}$~\msunyr. 
Moreover, recent 
analyses of the ellipsoidal variations of the M giant and the radial velocity 
data by Belczy\'nski \& Miko{\l}ajewska (1998) 
show that the system is a low 
mass binary system with stellar masses $\sim 0.7$~\msun\ for the giant and 
$\sim 1.2$~\msun\ for the hot component,
and solve practically all controversies about the nature of the hot component
and the physical causes of its eruptions. 

The recent outbursts of RS Oph, V3890 Sgr and V745 Sco, respectively,
in 1985, 1990 and 1989, indicate the outburst luminosity was $\ga L_{\rm{Edd}}$
for a 1.2~\msun\ white dwarf as predicted by TNR outburst models. In general, 
various outburst phenomena in these systems can be easily explained by the TNR 
models (see e.g.\ Anupama 1995). 
The amplitude and timescales of flickering detected during quiescence
in T CrB and RS Oph, 
are very similar to the flickering found in other cataclysmic variables,
and clearly point to the presence of an accreting white dwarf. 
The quiescent UV-optical spectrum of RS Oph on the other hand does not
exhibit the high ionization lines generally seen in the spectra of cataclysmic
variables, and occasionally also seen in T CrB. 
The high ionization emission lines are also absent in
the optical spectra of T CrB, RS Oph, V3890 Sgr and V745 Sco
analyzed in this paper.

The major problem for the TNR model is however the inconsistency
of the hot component's luminosity and effective temperature
(sec. 3.4; also Dobrzycka et al 1996a)
with standard massive white dwarf tracks in the HR diagram.
This seems to be an intrinsic feature of all four
recurrent novae discussed in our paper.
A possible explanation of this behaviour could be as follows.

The hot component in symbiotic systems is embedded in the red giant wind. 
In a majority of systems the hot component luminosity is high enough
to ionize a significant portion of the red giant wind giving rise
to both strong UV and strong optical emission lines. 
In particular, studies of samples of symbiotic systems in the radio
and optical range (e.g. Seaquist \& Taylor 1990;
Miko{\l}ajewska et al 1997), as well as numerical simulations 
of Raman scattered O\,{\sc vi} emission lines (Schmid 1996)
suggest that symbiotic systems have preferentially an ionization
geometry with an {\it X}-parameter (as defined by Seaquist et al 1984),
$X \sim 1$, which means that the `average shape' of the ionization
front does not differ significantly  from the plane between the two
stellar components.
However, in the case when the wind is very strong or/and the hot component
luminosity is relatively low, only a small region of the wind around the hot
companion is ionized and most of the emission emerging from this region
can be absorbed in the enveloping neutral wind. 
Shore \& Aufdenberg (1993), on the basis of 
an analysis of the ultraviolet spectra of several symbiotic systems, 
have shown that the emission lines could be severely 
affected by differential extinction due to absorption lines in the red
giant wind produced by neutral and 
singly ionized iron peak elements. The effect of this ``iron curtain'' is to 
lower the ultraviolet continuum temperatures and also suppress the emission 
lines at sufficiently high column densities ($\ge 5\times 10^{22}$~cm$^{-2}$). 
Shore \& Aufdenberg also point out that the absence of emission lines does not 
rule out the existence of an accretion disc around the hot component. 

Shore \& Aufdenberg (1993) and Shore et al (1996) have argued that 
the differential line absorption by the cool giant environment
can account for the shape and variability of UV spectra of RS Oph and T CrB.
The presence of Fe\,{\sc ii} emission lines in the optical region indicates that
line blanketing by the ``iron curtain'' is indeed possible. The ratio of the 
Fe\,{\sc ii} and the Ca\,{\sc ii} infrared triplet lines in RS Oph is similar to 
the ratios observed in active galactic nuclei (Joly 1989), and indicate electron 
densities of $\sim 10^{11}-10^{12}$~cm$^{-3}$ and column density 
$\ge 10^{23}$~cm$^{-2}$. The Balmer line ratios indicate high optical depth in 
the line emitting region. The O\,{\sc i} 8446/7774 line ratio in
RS Oph and V3890 Sgr clearly indicate enhancement of the 8446~\AA\ line by
Ly$\beta$ fluorescence. Further, these ratios also imply a mean Ly$\beta$
radiation density $\ga 1.2\times 10^3$~erg cm$^{-2}$ s$^{-1}$ sr$^{-1}$. This
implies the presence of a hot UV source hidden within the optically thick
wind envelope. The EUV luminosity of the hot component estimated using the
observed hydrogen and helium emission line fluxes is, as shown in section 3.4,
higher than the luminosities observed in normal stars of the estimated
spectral type. This is consistent with the hot source being embedded within the
wind envelope. The estimated spectral type of hot component is hence
not a true representation of the flux from the hot source and does not 
exclude the presence of a $\sim 10^5$~K white dwarf. Thus the temperature and 
radius estimated by Dobrzycka et al (1996a) for the hot component in RS Oph 
correspond to the pseudophotosphere of the wind envelope. 

The absence of high ionization lines can be accounted for by the absorption and 
softening by reradiation of all direct photons from the white dwarf+accretion 
disc. It may be noted that He\,{\sc ii} 4686~\AA\ line has been clearly detected 
in RS Oph in the spectrum of 1991 October (Iijima et al 1994),
indicating the presence of a $10^5$~K ionizing source. 
It is interesting to note that the visual magnitude was low ($\sim 11.8$ mag) 
during this period. We will return to this problem in the next section.
The 4686~\AA\ line was also
clearly present in the spectra of both V3890 Sgr and V745 Sco obtained 1--2 
years after the outburst.
The strength of this line decreased with time, with the re-establishment of
the envelope from the giant wind. We would like to point here that the
optical depths and electron densities and also the O\,{\sc i} line ratios in these
systems as indicated by their quiescence spectra are quite similar to what is 
observed in ``Fe\,{\sc ii}'' novae during the early phases of their outburst when the 
He\,{\sc ii} 4686~\AA\ line is absent (Anupama et al 1992; Kamath et al 1997). 

Finally, the absorption by the cool giant wind enveloping the hot
component can also account for the low quiescent X-ray luminosity of T CrB
and RS Oph (Selvelli et al 1992; Orio 1993).

\subsection{Causes of variability}

The most striking behaviour in the symbiotic RNe systems at quiescence is
the variability of the hot radiation flux, manifested as variability in the
UV and optical flux, and also in the fluxes in the emission lines. Selvelli et
al (1992) interpret the UV continuum and emission line variations in T CrB
as being due to changes in the photoionization source as a result of variable
mass transfer from the red giant. Shore \& Aufdenberg (1993) on the other hand
interpret the variations as effects of the environmental absorption. 
The lack of eclipses and absence of substantial changes in the 
spectrum of RS Oph rules out the brightening as being due to changes in the 
amount of intervening wind material.

Flickering detected in both T CrB and RS Oph can in principle provide some
information about fluctuations in the mass transfer/accretion rate.
In T CrB, the flickering is usually absent when the system is in a low 
state and present during the high state (see section 3.2). 
This indicates an increase in the mass
accretion rate during the optical brightening (the high state). 
The existing data on RS Oph do not
show any such variations, but as we note above (sec.3.2), the observations are
few and were all made when the star was relatively bright and blue.
We therefore cannot exclude fluctuations in the mass transfer
and consequently in the mass accretion rate also in this system. 
The flickering activity in T CrB was also investigated 
by Zamanov \& Bruch (1998). They identify the vicinity of a white
dwarf as the site of the flickering, and find that
the ratio of the flux of the flickering light source
and the quiet part of the primary remains constant
while the $U$ band flux varies considerably over long time scales.
The latter result suggests that in T CrB the mass is transferred
at the same rate from the secondary through the accretion
disk (a steady state configuration), and it seems to rule out
disk instabilities as the cause of the flickering.

As demonstrated by Shore \& Aufdenberg (1993), 
the changes in the UV spectrum from an absorption
spectrum to an emission line spectrum with no change in the total UV flux 
are due to changes in the line-of-sight column density 
through the wind envelope.
In particular, they found that $N_{\rm H} \sim 10^{22}-10^{23}\, \rm cm^{-2}$
is necessary to produce shell features in the UV spectrum and for the
emission lines to be suppressed.
Since in both T CrB and RS Oph the appearance of the shell features
and the lack of strong emission lines is generally not correlated with
the binary motion, to produce $N_{\rm H} \ga 10^{22}\, \rm cm^{-2}$, 
$\dot{M}_{\rm g}/v_{\rm g} \sim {\rm a\,\,few} \times 10^{-8} 
\rm M_{\sun}\,yr^{-1}/{\rm km\,s^{-1}}$ is required for an average 
$L_{\rm h} \sim 100$~\lsun.
It is interesting that this is roughly consistent with the accretion rate 
necessary to power the quiescent hot component luminosity in T CrB and RS Oph,
and to produce nova eruptions with the observed frequency.

Shore \& Aufdenberg (1993) also found that the mean column
density through the surrounding medium must be less than $5 \times 10^{21}
\rm cm^{-2}$ to observe strong emission lines. 
It is interesting that the only detection of the optical 
He\,{\sc ii} 4686 emission line in RS Oph coincided with notable decline
in the visual magnitude. This coincidence can be qualitatively
explained as follows. The decrease in the giant mass loss rate, and
in $N_{\rm H}$, results in disappearance of the absorbing species
from the line of sight, so the emission is less affected;
while the temporary decline in the mass accretion rate results
in decrease in the hot component luminosity.

Summarizing, we suggest that the changes in the UV spectrum during constant
$F_{\rm{UV}}$ phases, such as seen in T CrB, are due to absorption effects,
while the high and low states with corresponding changes in the luminosity
are due to fluctuations in the mass accretion rates possibly caused by a
variable mass loss from the red giant. The increase in the accretion rate during
the high states also gives rise to flickering. Fluctuation in mass accretion rate
can also cause variations in the absorption column density. The observed 
variability in these systems is thus more likely a combination of both effects,
i.e.\ fluctuations in the mass accretion rate as well as change in the absorption
column density.

Finally, the quiescent symbiotic recurrent
novae show striking similarities with CH Cyg, a highly variable 
symbiotic system. The hot component luminosity in CH Cyg varies
by a factor of $10^4$, between $\sim 0.1$
and $\sim 300$~\lsun\ (Miko{\l}ajewska 1994).
The brightening of the hot component is associated with the appearance
of flickering (e.g. Miko{\l}ajewski et al. 1990;
Miko{\l}ajewski \& Leedjaerv 1998), which suggests the system
is accretion-powered. The optical/UV spectrum of CH Cyg
during bright phases resembles the spectra of the symbiotic 
recurrent novae, in particular the shell-absorption emerges,
accompanied by H\,{\sc i}, He\,{\sc i} and Fe\,{\sc ii}
emission lines while the UV emission lines are weak or absent
(Miko{\l}ajewska et al 1988). The accreting component in this system is a
white dwarf (M{\"u}rset et al 1997; Ezuka et al 1998).

\section{Summary}

We present in this paper the behaviour of RNe with giant secondaries at
quiescence. Variability is detected in the emission line fluxes in correlation
with the continuum variations. The total UV flux is also found to be
variable in both RS Oph and T CrB. All these variations, which are related to the
radiation flux from the ionizing source, are caused by fluctuations in the 
mass accretion rates.

It is concluded that the hot source in these systems is an accreting white dwarf 
embedded in an optically thick envelope of wind from the giant. This envelope
absorbs the direct photons from the white dwarf. This picture can explain
the observed UV spectra, the X-ray luminosity, and the lack of high 
excitation lines in these systems. 

\begin{acknowledgements} This research was partly supported by
KBN Research Grant No. 2 P03D 021 012, and by exchange program
between Polish Academy of Sciences and Indian National Science
Academy (Program No. 7). We thank M. Friedjung and T.P. Prabhu for useful 
comments.
\end{acknowledgements}

{}


\begin{thebibliography}{}
\bibitem[]{}
Andrillat, Y., Houziaux, L., 1982, in the Nature of Symbiotic Stars, ed.\ 
Friedjung, M., Viotti, R. (Reidel: Dordrecht), p.57
\bibitem[]{}
Anupama, G.C. 1997, in Physical Processes in Symbiotic Binaries and Related 
Systems, ed. Miko{\l}ajewska, J. (Copernicus Foundation for Polish Astronomy: 
Warsaw), p.117
\bibitem[]{}
Anupama, G.C. 1995, in Cataclysmic Variables: Inter Class relations, eds.\
Bianchini, A., et al.\ (Kluwer Academic: Netherlands), p.49
\bibitem[]{}
Anupama, G.C., Duerbeck, H.W., Prabhu, T.P., Jain, S.K., 1992, A\&A, 263, 87
\bibitem[]{}
Anupama, G.C., Prabhu, T.P., 1989, JA\&A, 10, 237
\bibitem[]{}
Anupama, G.C., Prabhu, T.P., 1990, in IAU Colloquium No. 122, Physics of 
Classical Novae, ed. Cassatella, A., Viotti, R. (Springer-Verlag: Berlin), p. 
\bibitem[]{}
Anupama, G.C., Prabhu, T.P. 1991, MNRAS, 253, 605
\bibitem[]{}
Blair, W.P., Stencel, R.E., Feibelman, W.A., Michalitsianos, A.J., 1983, ApJS, 
53, 573
\bibitem[]{}
Belczy\'nski, K., Miko{\l}ajewska, J., 1998, MNRAS, 296, 77
\bibitem[]{}
Bianchini, A., Middleditch, J., 1976, Inf. Bull. Variable Stars, No.\, 1151 
\bibitem[]{}
Bruch, A., 1980, Inf. Bull. Variable Stars, No.\, 1805
\bibitem[]{}
Bruch, A., 1992, A\&A, 266, 237
\bibitem[]{}
Cassatella, A., Harris, A., Snijders, M.A.J., Hassall, B.J.M., 1985, ESA SP-236,
p.281
\bibitem[]{}
Canizzo, J.K., Kenyon, S.J., 1992, ApJ, 386, L17
\bibitem[]{}
Dobrzycka, D., Kenyon, S.J., 1994, AJ, 108, 2259
\bibitem[]{}
Dobrzycka, D., Kenyon, S.J., Proga, D., Miko{\l}ajewska, J., Wade, R.A., 1996a, 
AJ, 111, 2090
\bibitem[]{}
Dobrzycka, D., Kenyon, S.J., Milone, A.A.E., 1996b, AJ, 111, 414
\bibitem[]{}
Duerbeck, H.W., Seitter, W.C., 1990, in IAU Colloquium No. 122, Physics of 
Classical Novae, ed. Cassatella, A., Viotti, R. (Springer-Verlag: Berlin), p. 425
\bibitem[]{}
Evan, A., Callus, C.M., Albinson, J.S., Whitelock, P.A., Glass, I.S., Carter, B.,
Roberts, G., 1988, MNRAS, 234, 755
\bibitem[]{}
Garcia, M.R., 1986, AJ, 91, 1400
\bibitem[]{}
Harrison, T.E., Johnson, J.J., Spyromilio, J., 1993, AJ, 105, 320
\bibitem[]{}
Ianna, P.A., 1964, ApJ, 139, 780
\bibitem[]{}
Iijima, T., 1990, J. Am. Assoc. Variable Star Obs., 19, 28
\bibitem[]{}
Iijima, T., Strafella, F., Sabbadin, F., Bianchini, A., 1994, A\&A, 293, 919
\bibitem[]{}
Joly, M., 1989, A\&A, 208, 47
\bibitem[]{}
Kamath, U.S., Anupama, G.C., Ashok, N.M., Chandrasekhar, 1997, AJ, 114, 2671
\bibitem[]{}
Kastner, S.O., Bhatia, A.K., 1995, ApJ, 439, 346
\bibitem[]{}
Kato, M., 1990, ApJ, 355, 277
\bibitem[]{}
Kato, M., 1991, ApJ, 369, 471 
\bibitem[]{}
Kenyon, S.J., Garcia, M.R., 1986, AJ, 91, 125
\bibitem[]{}
Kenyon, S.J., Fernandez-Castro, T., 1987, AJ, 93, 938
\bibitem[]{}
Kenyon, S.J., Oliversen, N.A., Miko{\l}ajewska, J., Miko{\l}ajewski, M., Stencel, R.E.,
Garcia, M.R., Anderson, C.M., 1991, AJ, 101, 637
\bibitem[]{}
Kraft, R.P., 1958, ApJ, 127, 620
\bibitem[]{}
Lawerence, G.M., Ostriker, J.P., Hesser, J.E., 1967, ApJ, 148, L161
\bibitem[]{}
Lines, H.C., Lines, R.D., McFaul, T.G., 1988, AJ, 95, 1505
\bibitem[]{}
Luthardt, 1992, PASPC, 29,
\bibitem[]{}
Mason, K.O., C\'ordova, F.A., Bode, M.F., Barr, P., 1987, in RS Ophiuchi (1985) 
and the Recurrent Nova Phenomenon, ed. M.F. Bode (VNU Science Press, Utrecht), 
p.167
\bibitem[]{}
Miko{\l}ajewska, J., 1994, in Interacting Binary Stars, ed. A.W. Shafter,
ASP Conf. Ser., Vol. 56, p.374
\bibitem[]{}
Miko{\l}ajewska, J., Selvelli, P.L., Hack, M., 1988, A\&A, 198, 150
\bibitem[]{}
Miko{\l}ajewska, J., Acker, A., Stenholm, B., 1997, A\&A, 327, 191
\bibitem[]{}
Miko{\l}ajewski, M., Leedjaerv, L., 1998, IAU Circ. 
\bibitem[]{}
Miko{\l}ajewski, M., Miko{\l}ajewska, J., Tomov, T., Kulesza, B.,
Szczerba, R., Wikierski, B., 1990, Acta Astr., 40, 129
\bibitem[]{}
Miko{\l}ajewski, M., Tomov, T., Kolev, D., 1997, Inf. Bull. Variable Stars No. 4428
\bibitem[]{}
Orio, M., 1993, A\&A, 274, L41
\bibitem[]{}
Oskanian Jr., A.V., 1983, Inf. Bull. Variable Stars No.\ 2349
\bibitem[]{}
Raikova, D., Antov, A., 1986, Inf. Bull. Variable Stars No. 2960
\bibitem[]{}
Rosino, L. 1987, in RS Ophiuchi (1985) and the Recurrent Nova Phenomenon,
ed.\ Bode, M.F. (VNU Sci.\ Press: Utercht), p.1
\bibitem[]{}
Sanford, R.F., 1949, ApJ, 109, 81
\bibitem[]{}
Scott, A.D., Rawlings, J.M.C., Krautter, J., Evans, A., 1994, MNRAS, 268, 749
\bibitem[]{}
Schmid, H.M., 1996, MNRAS, 282, 511
\bibitem[]{}
Seaquist, E.R., Taylor, A.R., 1990, ApJ, 349, 313
\bibitem[]{}
Seaquist, E.R., Taylor, A.R., Button, S., 1984, ApJ, 284, 202
\bibitem[]{}
Sekiguchi, K., et al.\ 1990, MNRAS, 246, 78
\bibitem[]{}
Selvelli, P.L., Cassatella, A., Gilmozzi, R., 1992, ApJ, 393, 289
\bibitem[]{}
Shore, S.N., Aufdenberg, J., 1993, ApJ, 416, 355
\bibitem[]{}
Shore, S.N., Kenyon, S.J., Starrfield, S., Sonneborn, G., 1996, ApJ, 456, 717
\bibitem[]{}
Starrfield, S., Sparks, W.M., Shaviv, G., 1989, ApJ, 325, L35
\bibitem[]{}
Starrfield, S., Sparks, W.M., Truran, J.W., 1985, ApJ, 291, 136
\bibitem[]{}
Van Winckel, Duerbeck, H.W., Schwarz, H., 1993, A\&AS, 102, 401
\bibitem[]{}
Wade, R.A., 1982, AJ, 87, 1558
\bibitem[]{}
Walker, M.F., 1957, in Non-Stable Stars, IAU Symp. 3. ed. G.H. Herbig, p. 46
\bibitem[]{}
Walker, A.R., 1977, MNRAS, 179, 587
\bibitem[]{}
Webbink, R.F., Livio, M., Truran, J.W, Orio, M., 1987, ApJ, 314, 653
\bibitem[]{}
Williams, G., 1983, ApJS, 53, 523
\bibitem[]{}
Williams, R.E., Hamuy, M., Phillips, M.M., Heathcote, S.R., Wells, L., Nacarrete, M.,
1991, ApJ, 376, 721
\bibitem[]{}
Williams, R.E., Hamuy, M., Phillips, M.M., 1994, ApJS, 90, 297
\bibitem[]{}
Zamanov, R.K., Bruch, A., 1998, A\&A, 338, 988

\end{thebibliography}
\end{document}